# Orthogonal Frequency Division Multiplexing Continuous Variable Terahertz Quantum Key Distribution

Mingqi Zhang[1], and Kaveh Delfanazari[1]*

[1]*James Watt School of Engineering, University of Glasgow, Glasgow G12 8QQ, UK*

*Corresponding author: <u>kaveh.delfanazari@glasgow.ac.uk</u>

Date: 07 June 2025

*Abstract—* We propose a novel continuous-variable quantum key distribution (CVQKD) protocol that employs orthogonal frequency-division multiplexing (OFDM) in the terahertz (THz) band to enable high-throughput and secure quantum communication. By encoding quantum information across multiple subcarriers, the protocol enhances spectral efficiency and mitigates channel dispersion and atmospheric attenuation. We present a comprehensive security analysis under collective Gaussian attacks, considering both terrestrial free-space channels, accounting for humidity-induced absorption, and inter-satellite links, incorporating realistic intermodulation noise. Simulations show secret key rates (SKR) reaching ~72 bits per channel use in open-air conditions. While intermodulation noise imposes trade-offs, optimised modulation variance enables resilience and secure communication range. The maximum terrestrial quantum link extends up to 4.5 m due to atmospheric THz absorption, whereas inter-satellite links can support secure communication over distances exceeding 100 km, owing to minimal propagation channel losses in space. We evaluate the practical implementation of our protocol using recently developed on-chip coherent THz sources based on superconducting Josephson junctions. These compact, voltage-tunable emitters produce wideband coherent radiation, making them ideal candidates for integration in scalable quantum networks. By incorporating their characteristics into our simulations, we assess secure key generation under various environmental conditions. Our results show secure communication over distances up to 3 m in open air, and up to ~26 km in cryogenic or vacuum environments. This work advances the prospect of compact, high-capacity CVQKD systems for both terrestrial and space-based THz quantum communication.

## I. INTRODUCTION

Terahertz (THz) frequency region, located between microwaves and infrared, has caught the attention for its wide range of applications in quantum communication, biomedicine, material science, nondestructive evaluation, and sensing [1-3]. THz communication has been considered a key enabling technology of next-generation space-air-ground integrated networks, with wide operating bandwidth and high data rate capacity [4, 5]. However, THz hardware, including efficient sources, modulators, resonators, detectors, etc., is the core factor restricting the development of THz communication. Recent advancements in optical-to-THz conversion and compact solid-state THz sources offer promising solutions to bridge the THz gap [1], especially the superconducting coherent THz emitter technology, which offers wideband tunability [6-8], and on-chip frequency modulation with gigahertz bandwidth [9]. Such advanced hardware facilitates the development of THz communication and paves the way for novel applications in THz quantum technology. To enhance the security of THz communication, quantum key distribution (QKD), including both discrete variable (DV) and continuous variable (CV) approaches, is implemented to establish an unconditionally secure link between two parties [10]. CVQKD is particularly appealing for THz communication due to its high achievable rates, tolerance to high thermal noise, and compatibility with systems that do not require single-photon components [11, 12]. Recent research has demonstrated that THz CVQKD offers high data rate capacity and reliable wireless transmission in complex environments such as open-air, indoor, and inter-satellite scenarios while ensuring security through quantum cryptography [13, 14]. Furthermore, compared to single-carrier CVQKD protocols, utilising multi-carrier communication protocols has been shown to significantly extend the secure transmission distance and enhance the secret key rate [15, 16]. The multi-carrier protocol, which employs multiple transmitters and receivers, faces limitations in operating temperature within the mm-wave to low THz range and requires complex hardware implementation with on-chip integrated antennas [3]. However, by dividing a single high-speed data stream into multiple low-speed data streams and modulating them onto narrowband orthogonal subcarriers, multicarrier multiplexing (MCM) technology, particularly the orthogonal frequency-division multiplexing (OFDM) scheme, enables parallel key distribution using a single transmitter and receiver while efficiently utilising spectrum resources [17]. Although current research has demonstrated a promising increase in the secret key rate for THz MCM QKD, the underestimation of crosstalk and strong modulation noise introduced during imperfect processes, as well as the neglect of highly variable transmissivity for each subcarrier frequency in practical applications, may result in inaccurate performance estimations. To address this, we have introduced and analysed the impact of additional excess noise and incorporated the absorption coefficient for each subcarrier to more accurately evaluate the practical performance of the OFDM CVQKD scheme in the THz frequency range, bringing it closer to real-world applications. This paper is organised as follows: Section II presents the system model of the THz OFDM-based CVQKD protocol and introduces the modulation noise. Section III analyses the security of the scheme and computes the secret key rate. In Section IV, we provide simulation results for open-air, indoor, and inter-satellite environments. Section V discusses the protocol's performance concerning the latest advancements in THz hardware. Finally, we conclude the paper in Section VI.

## II. SYSTEM MODELING

*THz OFDM CVQKD protocol*

The THz OFDM-based CVQKD system scheme is illustrated in Fig.1. Alice generates a THz wave with an initial frequency $f_1$ and $N$ equally spaced frequency intervals $\Delta f$ orthogonal in the frequency domain from a THz emitter and then splits the wave into two signals with 90° phase difference. The two signals are fed into the local oscillator ports of two identical mixers, where they are combined with the I and Q signals from the OFDM generator. In the OFDM generator, Alice generates a stream of quantum random bits as vacuum states $|0\rangle$ and employs serial-to-parallel (S/P) conversion to transform the serial stream into $N$-independent parallel bit streams at the baseband. Each bit stream is modulated by a $q/p$-Gaussian modulator with a coherent state $|\alpha\rangle$, resulting in $|\alpha_k\rangle = |q_k + jp_k\rangle, (k = 1,2,...,N)$. The quadrature components $q_i$ and $p_i$ are drawn from two independent random variables $Q$ and $P$, both following a normal distribution $Q \sim P \sim N(0, V_{mod})$. The $N$-modulated states are then mapped onto their respective subcarrier $f_k=f_1+k\Delta f$ using an inverse Fourier transform (IFFT). The THz carriers are modulated by signals $I_s$ and $Q_s$ defined as [18]

$$I_s = \sum_{k=1}^{N} I_k \cos(2\pi f_k t) - \sum_{k=1}^{N} Q_k \sin(2\pi f_k t) \tag{1a}$$

$$Q_s = \sum_{k=1}^{N} I_k \sin(2\pi f_k t) + \sum_{k=1}^{N} Q_k \cos(2\pi f_k t) \tag{1b}$$

where $N$ is the number of subcarriers, $f_k=f_1+k\Delta f$ is the frequency of $k$-th subcarrier. The cyclic prefix (CP) is inserted into the signal to equalise the time domain at Bob's site, effectively mitigating inter-symbol and inter-channel interference during transmission [19]. Once the real and complex-valued outputs are converted to analogue signals, the encoded multi-carrier quantum state at Alice's side is represented as follows [18]

$$|X_{sig} + jP_{sig}\rangle = \otimes_{k=1}^{N} |X_k + jP_k + \Delta X_k + j\Delta P_k\rangle \tag{2}$$

where $X_k$ and $P_k$ are the modulated quadrature components for the $k$-th subcarrier, with modulation noise $\Delta X_k$ and $\Delta P_k$, respectively. The total variance of the $k$-th quantum mode prepared by Alice is $V_{A(k)}=V_{mod}+V_0$, where $V_0$ is the variance of the vacuum state given by [20]

$$V_0 = 2\bar{n} + 1 \tag{3}$$

with the vacuum shot noise unit (SNU) 1 and the mean thermal photon number

$$\bar{n} = \frac{1}{\exp\left(\frac{hf_k}{k_B T}\right) - 1} \tag{4}$$

where $h$ is Planck's constant, $f_k$ is the frequency, $k_B$ is the Boltzmann constant, and $T$ is the absolute temperature. The modulation noise in a practical OFDM-CVQKD protocol mainly arises from I/Q imbalance, which is caused by differences in signal amplitudes between the in-phase (I) and quadrature (Q) arms of the I/Q modulator. In addition, it is influenced by quadrature skew from the angular error $\theta$ of the 90° phase difference in the local oscillating signal, as well as intermodulation distortion. The quantum signal transmitted to Bob via an open-air, indoor or inter-satellite channel, with potential eavesdropping by Eve, can be written as [18]

$$E_{sig}(t) = 2A_{sig}e^{j2\pi f_A t}\left\{G_1 \sin\left[\sum_{k=1}^{N} \gamma_k \cos(2\pi f_k t + \varphi_k)\right] + jG_2 \sin\left[\sum_{k=1}^{N} \gamma_k \sin(2\pi f_k t + \varphi_k)\right]\right\} \tag{5}$$

$$\gamma_k = \mu_k \sqrt{I_k^2 + Q_k^2} \tag{6a}$$

$$\cos\varphi_k = \frac{I_k}{\sqrt{I_k^2 + Q_k^2}} \tag{6b}$$

$$G_1 = \frac{1 + \kappa_k e^{j\theta_k}}{2}, G_2 = \frac{1 + \kappa_k e^{-j\theta_k}}{2} \tag{6c}$$

where $A_{sig}$ is the amplitude of the signal, $f_A$ is the central frequency of the subcarriers, $G_1$ and $G_2$ are the I/Q imbalance factors, $0 \leq \kappa_k \leq 1$ is the transmitter gain imbalance, $0 \leq \theta_k \leq \frac{\pi}{2}$ is the quadrature skew and $\mu_k$ is the modulation index of the $k$-th subcarrier. At Bob's site, the received signal is combined with a locally generated signal in a coherent receiver to offset the carrier frequency. The signal is then digitised, the cyclic prefix is removed, and the serial quantum signal is converted into parallel form. A fast Fourier transform (FFT) is applied for demodulation to recover the signals. Finally, post-processing, including reconciliation and privacy amplification, is performed to finalise the secret keys.

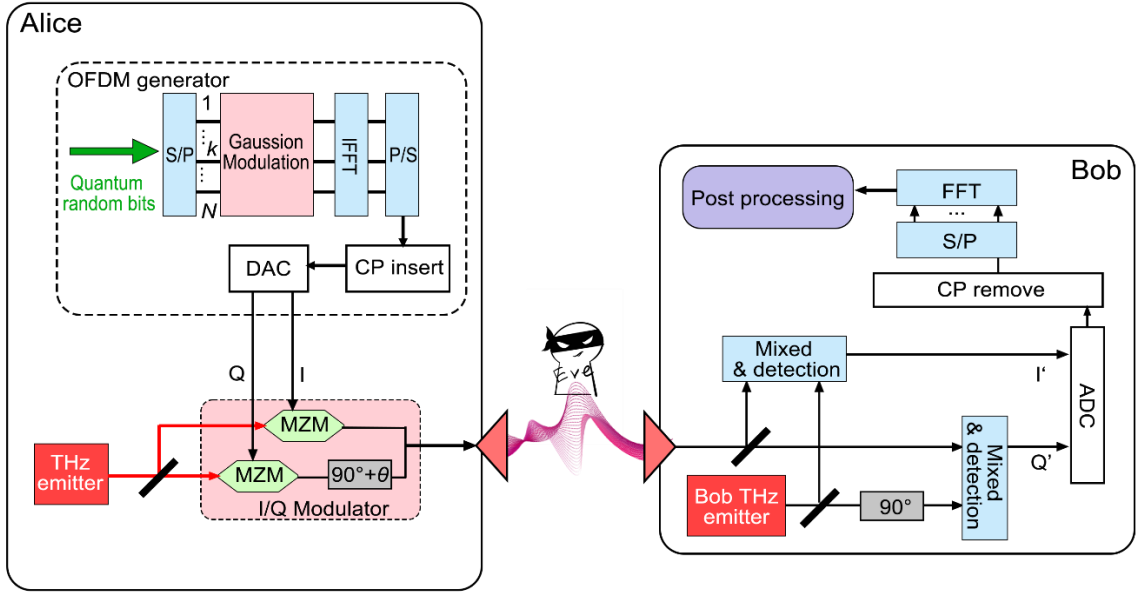

**Fig. 1.** The schematic of the THz wireless OFDM-CVQKD system. Components include: MZM (Mach-Zehnder modulator), S/P (serial to parallel conversion), IFFT (inverse fast Fourier transform), P/S (parallel to serial conversion), CP (cyclic prefix), DAC (digital to analogue conversion), ADC (analogue to digital conversion), and FFT (fast Fourier transform).

*Modulation noise*

Non-linear devices, such as amplifiers and mixers, can generate additional signals known as intermodulation products, including second-order distortion (2$f$) and third-order distortion (3$f$) from a single frequency, $f$. The amplitude of these distortions decreases with higher orders. When two frequencies pass through a nonlinear component, they mix, producing complex distortions across different orders. While most distortions occur far from the base frequencies $f_1$ and $f_2$, in-band signals from third-order distortion, such as 2$f_1$-$f_2$ and 2$f_2$-$f_1$, can be close to the fundamental signals and are difficult to eliminate using filters. Third-order distortions may overlap with the fundamental signals for three input frequencies with the same gap $\Delta f$, such as $f_3 = f_2 + \Delta f = f_1 + 2\Delta f$. For example, 2$f_2$- $f_3$= $f_1$ and 2$f_2$- $f_1$= $f_3$. In OFDM-based multi-carrier modulation, multiple subcarrier signals with a fixed frequency interval transmit information in parallel. We assume the frequency gap between two subcarriers is $\Delta f$ and use $N$ subcarriers in the system from the fundamental frequency $f$. We consider the worst-case scenario where the third-order distortions of $f_m, f_n, f_l$ coincide with the signal $f_k$, when $m, n, l, k \in [1,N]$.

To estimate the modulation noise in Eq. (4), previous work [18] applies the Jacobi-Anger expansion and uses Bessel functions to express the quadrature $X_{sig(k)}$ of the $k$-th subcarrier as

$$X_{sig(k)} = X_k + \Delta X_k = 2A_{sig}\mu_k I_k + \Delta X_{k1} + \Delta X_{k2} + \Delta X_{k3} \tag{7}$$

Here, $X_k$ represents the $k$-th subcarrier component without intermodulation distortion or I/Q imbalance ($\kappa_k = 1, \theta_k = 0$), and the extra modulation noise $\Delta X_k$ consists of three parts [18]:
1. The modulation noise from the I/Q imbalance

$$\Delta X_{k1} = A_{sig}\mu_k[(\kappa_k cos\theta_k - 1)I_k + \kappa_k sin\theta_k Q_k] \tag{8a}$$

2. The modulation noise from the third-order intermodulation with $f_k=2f_m-f_n$ and $f_k=2f_m+f_n$, with condition $k=2m-n$ and $k=2m+n+2f/\Delta f$, respectively

$$\Delta X_{k2} = \frac{1}{4} A_{sig} \mu_m^2 \mu_n \left\{ (1 + \kappa_k \cos\theta_k) \begin{Bmatrix} [M_1(N,k) + M_2(N,k)]Q_m^2 I_n \\ +2[M_1(N,k) - M_2(N,k)]I_m Q_m Q_n \\ -[M_1(N,k) + M_2(N,k)]I_m^2 I_n \end{Bmatrix} + \kappa_k \sin\theta_k \begin{Bmatrix} [M_1(N,k) + M_2(N,k)]I_m^2 Q_n \\ +2[M_1(N,k) - M_2(N,k)]I_m Q_m I_n \\ -[M_1(N,k) + M_2(N,k)]Q_m^2 Q_n \end{Bmatrix} \right\} \quad (8b)$$

3. The modulation noise from the third-order intermodulation with $f_k=f_m+f_n+f_l$, $f_k=f_m+f_n-f_l$ and $f_k=f_m-f_n-f_l$, with condition $k=m+n+l+2f/\Delta f$, $k=m+n-l$ and $k=m-n-l-2f/\Delta f$, respectively

$$\Delta X_{k3} = \frac{1}{4} A_{sig} \mu_m \mu_n \mu_l \left\{ (1 + \kappa_k \cos\theta_k) \begin{Bmatrix} [W_1(N,k) - W_2(N,k) + W_3(N,k)]I_m Q_n Q_l \\ +[W_1(N,k) - W_2(N,k) - W_3(N,k)]Q_m I_n Q_l \\ +[W_1(N,k) + W_2(N,k) - W_3(N,k)]Q_m Q_n I_l \\ -[W_1(N,k) + W_2(N,k) + W_3(N,k)]I_m I_n I_l \end{Bmatrix} + \kappa_k \sin\theta_k \begin{Bmatrix} [W_1(N,k) + W_2(N,k) - W_3(N,k)]I_m I_n Q_l \\ +[W_1(N,k) - W_2(N,k) - W_3(N,k)]I_m Q_n I_l \\ +[W_1(N,k) - W_2(N,k) + W_3(N,k)]Q_m I_n I_l \\ -[W_1(N,k) + W_2(N,k) + W_3(N,k)]Q_m Q_n Q_l \end{Bmatrix} \right\} \quad (8c)$$

Here, $M_1(N,k)$ and $M_2(N,k)$ are the combinatorial numbers of the conditions $k=2m-n$ and $k=2m+n+2f/\Delta f$, respectively. Similarly, $W_1(N,k)$, $W_2(N,k)$, and $W_3(N,k)$ are the combinatorial numbers of the conditions $k=m+n+l+2f/\Delta f$, $k=m+n-l$, and $k=m-n-l-2f/\Delta f$, respectively. The modulation indices $\mu_k$, $\mu_m$, $\mu_n$, and $\mu_l$ correspond to the $k$, $m$, $n$, $l$-th subcarriers and are approximately equal to one another. The quadrature $I_g$ and $Q_g$ ($g=k, m, n, l$) are mutually random and independent, sampled from the distribution $Q \sim P \sim N(0, V_{mod})$ with variance $\langle I_g^2 \rangle = \langle Q_g^2 \rangle = \sigma_1^2 = V_{mod}$ and forth moments $\langle I_g^4 \rangle = \langle Q_g^4 \rangle = \sigma_2^2 = 2\sigma_1^4$ [18]. Therefore, the variance of $\Delta X_k$ can be described as [18]

$$\langle \Delta X_k^2 \rangle = \langle \Delta X_{k1}^2 \rangle + \langle \Delta X_{k2}^2 \rangle + \langle \Delta X_{k3}^2 \rangle \quad (9)$$

$$\langle \Delta X_{k1}^2 \rangle = A_{sig}^2 \mu_k^2 \sigma_1^2 (\kappa_k^2 + 1 - 2\kappa_k \cos\theta_k) \quad (10a)$$

$$\langle \Delta X_{k2}^2 \rangle = \frac{A_{sig}^2 \mu_m^4 \mu_n^2}{8} (1 + 2\kappa_k \cos\theta_k + \kappa_k^2) \begin{Bmatrix} [M_1(N,k) + M_2(N,k)]^2 \sigma_2^2 \sigma_1^2 \\ +2[M_1(N,k) - M_2(N,k)]^2 \sigma_1^6 \end{Bmatrix} \quad (10b)$$

$$\langle \Delta X_{k3}^2 \rangle = \frac{A_{sig}^2 \mu_m^2 \mu_n^2 \mu_l^2}{4} (1 + 2\kappa_k \cos\theta_k + \kappa_k^2)[W_1^2(N,k) + W_2^2(N,k) + W_3^2(N,k)] \sigma_1^6 \quad (10c)$$

Based on Eq. (9) and (10), the modulation noise of the k-th subcarrier can be expressed as

$$\varepsilon_{mod}(k) = \langle \Delta X_k^2 \rangle = A_{sig}^2 \mu_k^2 \sigma_1^2 (\kappa_k^2 + 1 - 2\kappa_k \cos\theta_k) + \frac{A_{sig}^2 \mu_k^6 \sigma_1^2}{8} (1 + 2\kappa_k \cos\theta_k + \kappa_k^2) \left\{ \begin{matrix} [M_1(N,k) + M_2(N,k)]^2 \sigma_2^2 \\ +2\{[M_1(N,k) - M_2(N,k)]^2 \\ +W_1^2(N,k) + W_2^2(N,k) + W_3^2(N,k)\} \sigma_1^4 \end{matrix} \right\} \quad (11)$$

By choosing a lower modulation index $\mu_k=0.01$, an amplitude $A_{sig}=1$, a gain imbalance $\kappa_k=0.98$ and a quadrature $\theta_k=\pi/50$, we show the modulation noise $\varepsilon_{mod}(k)$ of the $k$-th subcarrier in Fig.2 (a) under different total carrier numbers $N$. The highlighted points on each line represent the maximum modulation noise in the $N$-subcarrier system, defined as $\varepsilon_{mod}(k_{worst})$. It is evident that the middle subcarriers experience higher modulation noise, and the modulation noise level across all subcarriers increases as the total carrier number grows. Figure 2 (b) further demonstrates the exponential growth of the $\varepsilon_{mod}(k_{worst})$ as more subcarriers are used in the protocol. The maximum modulation noise grows significantly from $N=10$ and reaches 27 SNU at $N=120$, making it too large to ignore. Therefore, it is essential to account for the modulation noise at Alice's site.

The modulation noise is considered as part of the excess noise at Alice's side [21]. In multi-carrier CVQKD, the modulation noise $\varepsilon_{mod}(k)$ represents extra noise added to the excess noise $\varepsilon_{single}$ of a single-carrier CVQKD scheme. Consequently, the total excess noise of the $k$-th subcarrier can be written as [18]

$$\varepsilon_{multi}(k) = \varepsilon_{single}(k) + \varepsilon_{mod}(k) \quad (12)$$

where, $\varepsilon_{single}(k)$ represents the loss-induced vacuum noise associated with the transmissivity $T_{ch}(k)$ of the $k$-th subcarrier frequency [22]

$$\varepsilon_{single}(k) = \frac{1 - T_{ch}(k)}{T_{ch}(k)} \quad (13)$$

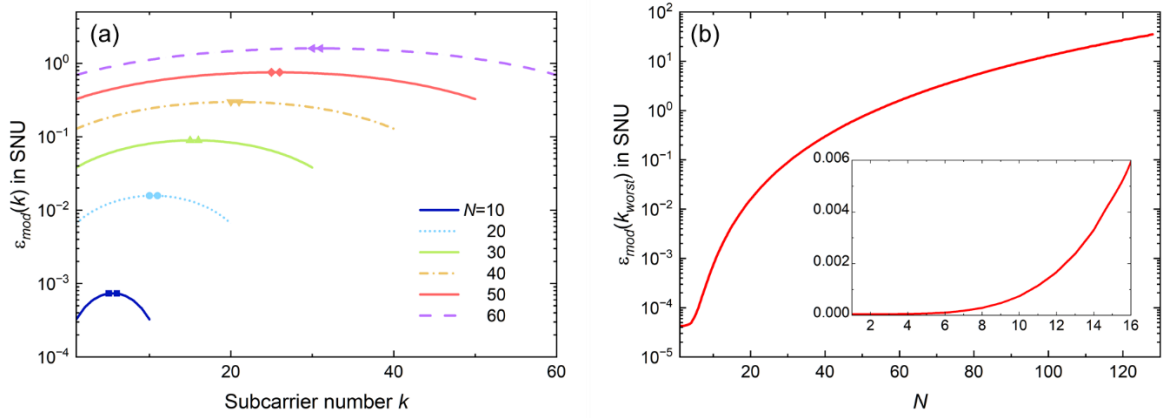

**Fig. 2.** (a) Simulated modulation noise $\varepsilon_{mod}(k)$ of the $k$-th subcarrier with $N$ ranging from 10 to 60. The solid points on each line indicate the maximum modulation noise achieved by the $k_{worst}$-th subcarrier among the $N$ subcarriers. (b) Worst modulation noise $\varepsilon_{mod}(k_{worst})$ of the $k_{worst}$-th subcarrier across $N$ subcarriers. The inset is the zoomed-in view for smaller values of $N$.

## III. SECRET KEY RATE

To evaluate the security and achievable distance of the THz OFDM CVQKD scheme, we compute the secret key rate (SKR) $R$ using reverse reconciliation in this section. The multi-carrier scheme is considered an $N$-independent orthogonal transmission occurring simultaneously [17]. Hence, the secret key rate of the OFDM CVQKD system is the sum of the secret key rates of the $N$ subcarriers given by [15]

$$R_{OFDM} = \sum_{k=1}^{N} R_k \tag{14}$$

For the $k$-th subcarrier, the secret key rate (bits/channel use) quantifies the extra private information shared by Alice and Bob compared to the information accessible to Eve. It is expressed as

$$R_k = \beta I_{AB_k} - I_{BE_k}$$

where $I_{AB_k}$ is the Shannon mutual information between Alice and Bob, $I_{BE_k}$ is the accessible information of Eve bounded by the Holevo information, and $\beta$=1 represents the efficiency of reverse reconciliation. The mutual information is defined as [20]

$$I_{AB_k} = \frac{1}{2} \log_2 \frac{V_b}{V_{b|a}} \tag{15}$$

where $V_b$ is the variance of Bob's mode and is given by

$$V_b = \eta T_{ch}(V_A + \varepsilon_{multi}) + \eta(1 - T_{ch})W + (1 - \eta)S \tag{16}$$

where the detection efficiency $\eta$=0.1 and the trusted thermal noise at Bob's site is $S$=1 [20]. The conditional variance is similar to Eq. (16) but with the Gaussian modulation variance $V_{mod}$=0, yielding: $V_{b|a} = \eta T_{ch}(V_0 + \varepsilon_{multi}) + \eta(1 - T_{ch})W + (1 - \eta)S$.

We assume Eve employs a collective entangling cloner attack, the most powerful and practical attack against Gaussian protocol in the channel. The thermal noise $W$ is injected by Eve using her two-mode squeezed state. Eve's information about Bob's measurement, derived from the Holevo bound is $I_{BE_k} = S_{E_k} - S_{E|B_k}$ with the von Neumann entropy

$$S = \sum h(x) \tag{16a}$$

$$h(x) = \frac{x+1}{2} \log_2 \frac{x+1}{2} - \frac{x-1}{2} \log_2 \frac{x-1}{2} \tag{16b}$$

where $x$ represents the symplectic eigenvalues of the covariance matrix. After transmission, the covariance matrix for Eve's information in the $k$-th subcarrier is given by

$$\mathbf{V}_{E_k} = \begin{bmatrix} W\mathbf{I} & \sqrt{T_{ch_k}(W^2-1)}\mathbf{Z} \\ \sqrt{T_{ch_k}(W^2-1)}\mathbf{Z} & [(1-T_{ch_k})V_{A_k} + T_{ch_k}W]\mathbf{I} \end{bmatrix} \tag{17}$$

where $\mathbf{I}$=diag(1,1), $\mathbf{Z}$=diag(1,-1). We set $a = W$, $b = (1-T_{ch_k})V_{A_k} + T_{ch_k}W$, and $c = \sqrt{T_{ch_k}(W^2-1)}$. The symplectic eigenvalues $v_1$ and $v_2$ for the two-mode covariance matrix $\mathbf{V}_{E_k}$ can be calculated from $v$=|$i\mathbf{\Omega V}_{E_k}$|, with $\mathbf{\Omega}$ defining the symplectic

form given by $\Omega := \bigoplus_{k=1}^{n} \begin{pmatrix} 0 & 1 \\ -1 & 0 \end{pmatrix}$ and $\oplus$ is the direct sum denoting adding matrices on the block diagonal [23]. The symplectic eigenvalues can be described as

$$v_{1,2} = \frac{1}{2}[z \pm (a-b)] \tag{18}$$

where $z = \sqrt{(a+b)^2 - 4c^2}$ [23].
The conditional covariance matrix for Eve's information in the $k$-th subcarrier can be calculated from

$$\mathbf{V}_{E|B_k} = \mathbf{V}_{E_k} - V_{B_k}^{-1} \mathbf{D} \mathbf{\Pi} \mathbf{D}^{\mathrm{T}} \tag{19}$$

with

$$\mathbf{D} = \begin{pmatrix} X\mathbf{I} \\ Y\mathbf{Z} \end{pmatrix}, \mathbf{\Pi} := \begin{pmatrix} 1 & 0 \\ 0 & 0 \end{pmatrix} \tag{20}$$

where $X = \sqrt{\eta T_{ch_k}(1 - T_{ch_k})(W - V_{A_k})}$ and $Y = \sqrt{\eta(1 - T_{ch_k})(W^2 - 1)}$ [16, 23, 24]. Using Eq. (19), Eve's conditional covariance matrix can be written as

$$\mathbf{V}_{E|B_k} = \begin{pmatrix} \mathbf{A} & \mathbf{C} \\ \mathbf{C}^{\mathrm{T}} & \mathbf{B} \end{pmatrix} \tag{21}$$

where

$$\mathbf{A} = \begin{pmatrix} b - \frac{X^2}{V_{B_k}} & 0 \\ 0 & b \end{pmatrix}, \mathbf{B} = \begin{pmatrix} a - \frac{Y^2}{V_{B_k}} & 0 \\ 0 & a \end{pmatrix}, \mathbf{C} = \begin{pmatrix} c - \frac{XY}{V_{B_k}} & 0 \\ 0 & -c \end{pmatrix} \tag{22}$$

The symplectic eigenvalues $v_3$ and $v_4$ of the matrix in Eq.(21) are given by [23, 24]

$$v_{3,4} = \sqrt{\frac{\Delta \pm \sqrt{\Delta^2 - 4\det \mathbf{V}_{E|B_k}}}{2}} \tag{23}$$

where $\Delta := \det \mathbf{A} + \det \mathbf{B} + 2\det \mathbf{C}$ and det denote the determinant. Finally, the mutual information between Eve and Bob using Eq.(16), (18), and (23) is

$$I_{BE_k} = S_{E_k} - S_{E|B_k} = h(v_1) + h(v_2) - h(v_3) - h(v_4) \tag{24}$$

IV. RESULTS

*Open-air channel*
Water vapour, oxygen, and other atmospheric species can absorb electromagnetic waves across a wide range, from radio waves to the THz region, and even cause opacity in certain frequency bands. This absorption significantly degrades signal propagation quality through the channel. In the THz frequency bands, water vapour is the dominant absorption component in both open-air and indoor environments. Therefore, obtaining an accurate water vapour absorption spectrum is crucial for identifying atmospheric windows between strong absorption lines, enabling efficient THz communication.

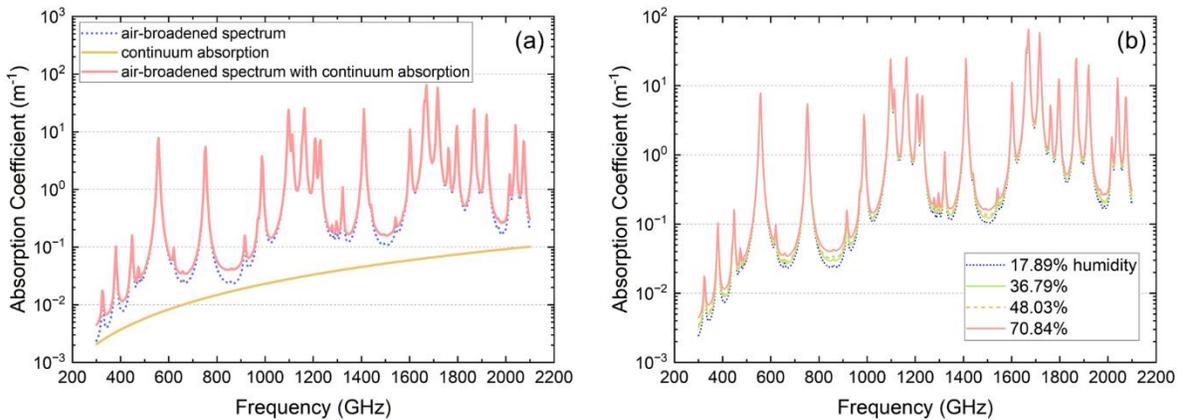

**Fig. 3.** (a) Simulated air-broadened spectrum and continuum absorption at 70.84% relative humidity and room temperature. (b) Simulated air-broadened spectrum with continuum absorption for relative humidity levels of 17.89%, 36.79%, 48.03%, and 70.84% at room temperature.

Water vapour absorption consists of two components: resonant line absorption, which can be modelled as the sum of all resonant lines using spectroscopic database data and line shape functions, and continuum absorption, which is experimentally observed as the difference between the measured total absorption and the resonant line absorption [25]. Using the method described in [26], we simulated the absorption coefficient $\alpha_f = \alpha_l + \alpha_c$ from 300 GHz to 2.1 THz. This includes the air-broadened spectrum $\alpha_l$ and the continuum absorption $\alpha_c$, as shown in Fig.3 (a). The blue dashed line shows the air-broadened spectrum due to all the resonant absorption lines, calculated using parameters from HITRAN 2020 [27], under a relative humidity of 70.84% with 15.23 Torr of water vapour and 746.77 Torr of air. The yellow solid line represents the continuum absorption spectrum, which accounts for deviations between calculation and observation. The pink solid line shows the accurately modelled absorption coefficient $\alpha_f$. In Fig.3 (b), we compare the absorption coefficient at humidity levels of 17.89%, 36.79%, 48.03%, and 70.84% corresponding to water vapour pressures of 3.60, 8.04, 10.05, 15.23 Torr, respectively. The results indicate several atmospheric windows with low absorption coefficients, making them suitable for applications. The relative humidity strongly influences these windows.

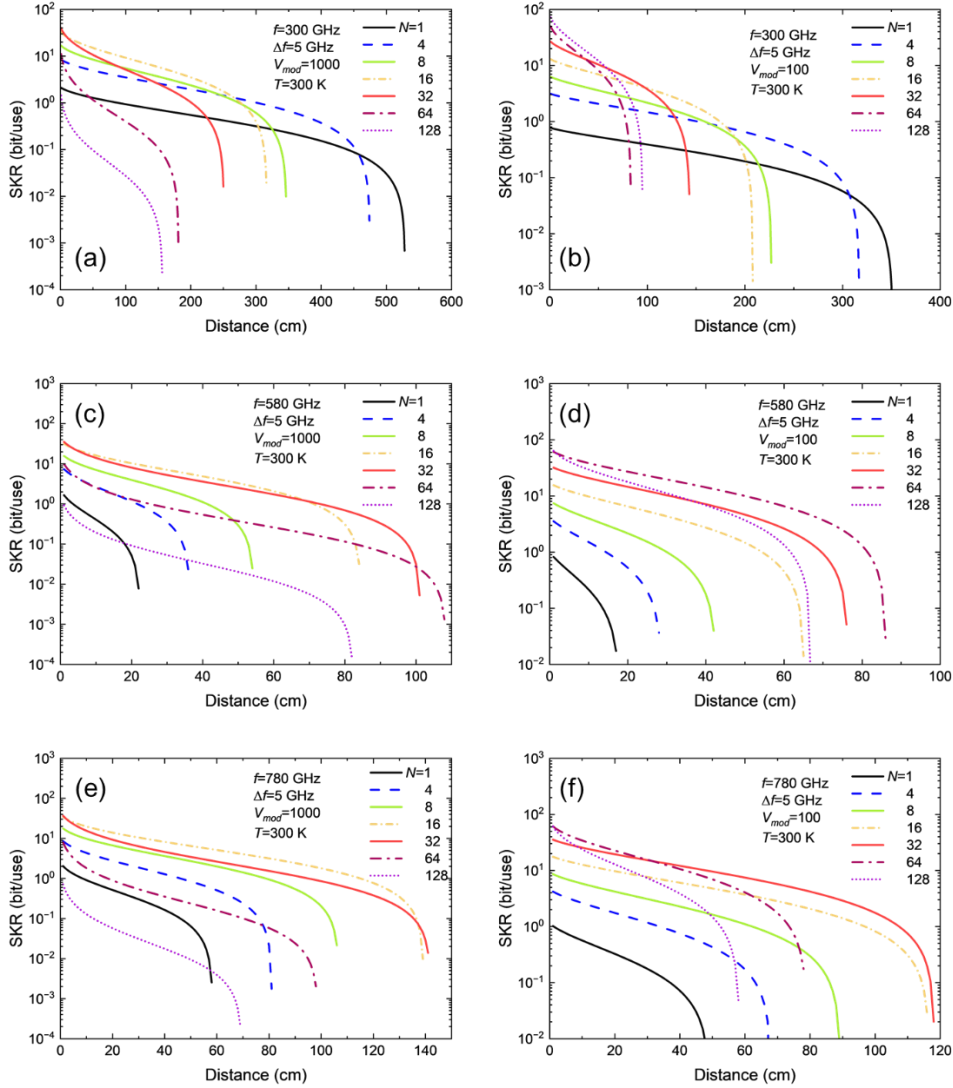

**Fig. 4.** SKR of THz-OFDM-CVQKD as a function of transmission distance using homodyne detection for different frequencies and modulation variances. (a) $f = 300$ GHz, $\Delta f = 5$ GHz, $V_{mod} = 1000$; (b) $f = 300$ GHz, $\Delta f = 5$ GHz, $V_{mod} = 100$; (c) $f = 580$ GHz, $\Delta f = 5$ GHz, $V_{mod} = 1000$; (d) $f = 580$ GHz, $\Delta f = 5$ GHz, $V_{mod} = 100$; (e) $f = 780$ GHz, $\Delta f = 5$ GHz, $V_{mod} = 1000$; (f) $f = 780$ GHz, $\Delta f = 5$ GHz, $V_{mod} = 1000$, all evaluated at 300K temperature, 70.84% relative humidity. The number of subcarriers $N$ varies as $N=1, 4, 8, 16, 32,$ and 128, considering the worst-case modulation noise $\varepsilon_{mod}(k_{worst})$ for each $N$.

The transmissivity of the open-air link depends on the absorption coefficient $\alpha_f$ at frequency $f$ and transmission distance $d$, given by the Beer-Lambert Law [26]:

$$T_{ch} = \exp(-\alpha_f d) \tag{7}$$

We select three main frequency regions that cover absorption peaks and windows, based on recent hardware advances [28-31]. Figure 4 shows the simulation results for the secure distance from fundamental frequencies $f$=300 GHz, $f$=580 GHz, and $f$=780

GHz, with a bandwidth of $\Delta f$=5 GHz at 300 K. Different modulation variances, specifically 1000 and 100, were considered for each frequency range. The increase in subcarriers from $N$=1 to $N$=32 in Fig.4 (a), (c), and (e) enhances the SKR for communication. Previous studies, which did not account for modulation noise, suggested that the SKR increases linearly with the number of subcarriers in the scheme [16]. However, for all three communication bands as shown in Fig. 4 (a), (c), and (e), the SKR of $N$=64 is much lower than $N$=16, and for $N$=128, it is even smaller than that of a single carrier.

Therefore, the impact of high modulation noise between multiple subcarriers becomes significant, causing a substantial reduction in the SKR from $N$=32 to $N$=128. While the secure distance decreases for all frequencies and subcarrier counts when the modulation variance is reduced to $V_{mod}$=100, as shown in Fig. 3 (b), (d), and (f), the system becomes more robust against modulation noise. This is evidenced by the relatively stable SKR for $N$=32 to $N$=128 under lower $V_{mod}$. However, even with $V_{mod}$=100, the increase in SKR is hindered at $N$=128 due to the effects of modulation noise, especially for higher frequencies in Fig. 3 (d) and (f) compared with $f$=300 GHz in Fig.3 (b). The highest SKR can be reached by $N$=128, $V_{mod}$=100 in Fig. 3(b) as ~72 bit/channel use, while the longest distance with multicarrier can be reached > 4.5 m in Fig. 3 (a) by $N$=4, $V_{mod}$=1000.

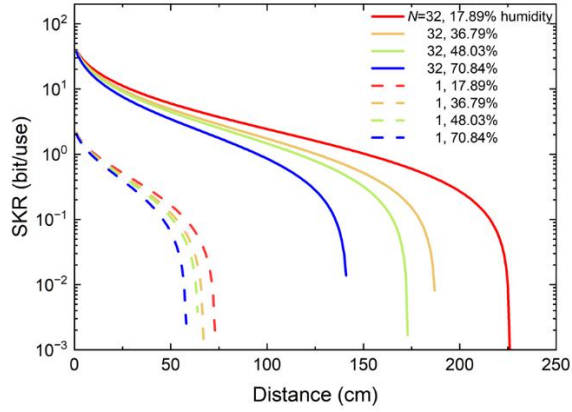

**Fig. 5.** SKR as a function of transmission distance using homodyne detection at $f = 580$ GHz, with $\Delta f = 5$ GHz evaluated under relative humidities of 17.89%, 36.79%, 48.03%, and 70.84%. The dashed lines represent the single-carrier configuration ($N$=1), and the solid lines correspond to the multi-carrier case ($N$=32). The modulation variance is $V_{mod}$=1000, and the worst modulation noise $\varepsilon_{mod}(k_{worst})$ is considered for each $N$.

The initial SKRs for $N$<32 with $V_{mod}$=100 are slightly lower compared to $V_{mod}$=1000 for each frequency range. Therefore, while a higher $V_{mod}$ enables longer distances and higher SKRs, it is highly sensitive to modulation noise. Conversely, a lower $V_{mod}$ reduces the transmission distance and SKR but enhances robustness against modulation noise across multiple carriers. The cutoff distance is related to $N$ and highly depends on the THz absorption coefficient at the chosen operating frequency. Unlike Figs 4 (c)-(f), where higher subcarrier counts can extend the secure distance, a single carrier achieves the greatest secure distance in the carrier frequency range starting from $f$=300 GHz in Fig.4 (a) and (b). This is because $f$=300 GHz corresponds to the lowest absorption point in the exponential increase of the absorption coefficient. Consequently, each additional subcarrier beyond $f$ encounters higher absorption, reducing the secure distance for its transmitted information compared to 300 GHz. The same reason causes the non-linear relationship of large $N$ and distance, that the significant absorption peak in the carrier frequency range leads to a short secure distance for the subcarrier located on the peak. Drawing on the humidity-induced absorption data from Fig. 3(b), we analysed the OFDM CVQKD protocol's performance at $f$=580 GHz with $V_{mod}$=1000 and 300 K under varying relative humidity in Fig. 5. For both single-carrier ($N$=1) and multi-carrier ($N$=32) configurations, reducing the humidity extends the transmission distance without affecting the initial secret key rate. By using 32 subcarriers spanning 580 GHz to 735 GHz in 5 GHz increments, a secure distance of 2.25m could be achieved at 17.89% relative humidity- almost triple that of a single carrier. Additionally, the secret key rate increases by over 15-fold. In practice, the choice of operating frequencies (based on the absorption coefficient) and the balance among modulation noise, achievable SKR, and secure distance must align with specific application requirements. For instance, short-range indoor wireless links between quantum computers may only need a few meters of secure transmission. Under these circumstances, using multiple subcarriers and maintaining low humidity levels can be highly beneficial.

*THz OFDM CVQKD Cryogenic links*

Employing wireless communication within a quantum computer or cryostat can free valuable space by eliminating the need for bulky transmission lines, which often restrict the number of quantum circuits. In a cryogenic environment, the transmission link behaves similarly to inter-satellite channels, typically modelled as diffraction-limited, with transmissivity given by [32]

$$T_{ch} = 1 - \exp\left(\frac{2r_a^2}{\omega^2(d)}\right) \tag{8}$$

where $\omega(d)$ is the beam radius at distance $d$, and $r_a$ is the receiver aperture radius. We set the beam-waist radius equal to the receiver aperture radius, i.e., $\omega_0 = r_a = 10$ cm. Therefore, the beam radius is calculated as

$$\omega(d) = \omega_0 \times \sqrt{1 + \left(\frac{\lambda d}{\pi \omega_0^2}\right)^2} \tag{9}$$

where $\lambda$ is the wavelength of the sub-carrier beam [32].

Figure 6 shows the simulation results of the SKR in a vacuum channel at 30 K with $f$=780 GHz and $\Delta f$=5 GHz, for (a) $V_{mod}$=100 and (b) $V_{mod}$=1000. Unlike open-air links, the secure distance increases linearly with the number of subcarriers due to the absence of atmospheric absorption. For a minimum SKR of $10^{-5}$ bit/channel use, a 128-sub-carrier system with $V_{mod}$=1000 achieves a link distance of up to 116 km. However, in Fig. 6(b), the higher modulation variance leads to more pronounced SKR decay starting from $N$=64, as the modulation noise between subcarriers begins to significantly impact the SKR. In comparison, Fig. 6(a), with a lower modulation variance, shows only a slight decrease in SKR even at $N$=128. Thus, in both open-air and cryogenic vacuum channels, employing a lower modulation variance enhances robustness against inter-subcarrier modulation noise.

In Fig. 7, we examine a 32-subcarrier system operating across various frequencies. As the starting frequency increases from $f$=600 GHz to $f$=8.1 THz, the secure distance largely improves. It reaches over 200 km at $f$=8.1 THz with $V_{mod}$=100 at 30 K. Additionally, the SKR experiences a modest enhancement at these higher frequencies. Thus, to extend secure distances and boost SKR, it is essential to employ high frequencies along with an optimal subcarrier count that effectively mitigates modulation noise.

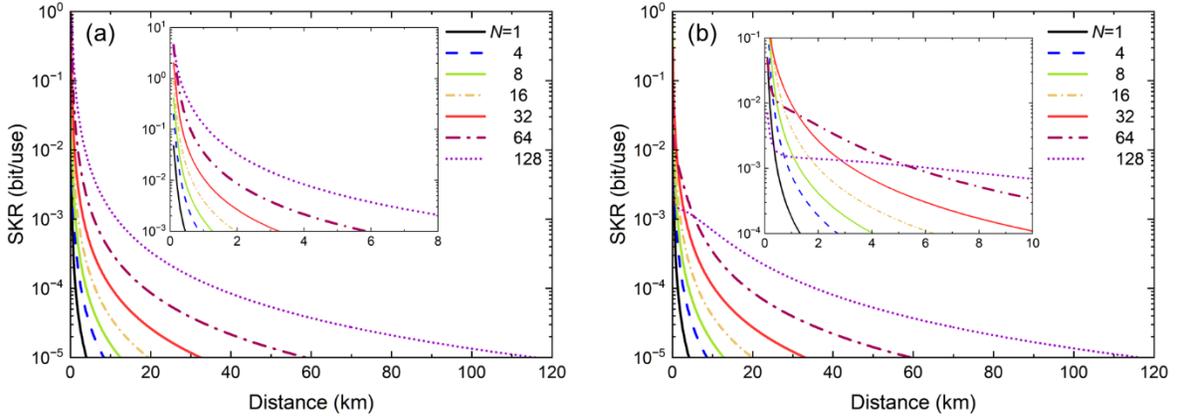

**Fig. 6.** SKR as a function of transmission distance in vacuum channel at 30K using homodyne detection with $f = 780$ GHz, and $\Delta f = 5$ GHz. (a) $V_{mod}$=100. (b) $V_{mod}$=1000. The analysis accounts for the worst-case modulation noise $\varepsilon_{mod}(k_{worst})$ for each subcarrier number $N$.

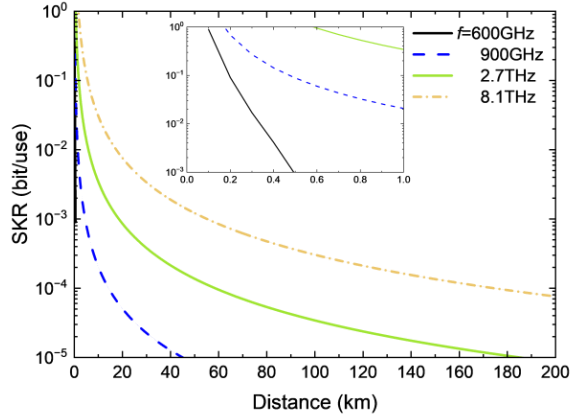

**Fig. 7.** SKR as a function of transmission distance in a vacuum channel at 30K using homodyne detection. The starting frequencies are $f$=600 GHz, 900 GHz, 2.7 THz, and 8.1 THz, with $\Delta f = 5$ GHz. The system employs 32 subcarriers ($N$=32) and a modulation variance of $V_{mod}$=100. The worst-case modulation noise $\varepsilon_{mod}(k_{worst})$ has been considered.

## V. PROPOSAL FOR THz QKD WITH SUPERCONDUCTING COHERENT PHOTON SOURCES

To achieve THz-range quantum communication with multi-subcarriers, a tunable THz source with a broad radiation bandwidth is essential. High-temperature superconducting coherent THz emitters, integrating large arrays of phase-synchronised Josephson junctions, generate powerful and wideband THz waves at temperatures near or above 77 K (liquid nitrogen boiling point), demonstrating significant potential for quantum applications [33-35]. Without requiring a complex and bulky optical setup, these THz emitters operate with a DC voltage, generating 0.1–11 THz waves at the µW power scale [7, 36, 37]. Moreover, these emitters can generate continuously modulated carrier waves with bandwidths up to 40 GHz, centred at 840 GHz [9, 38]. While a large gap remains between current technological advancements and the practical deployment of a THz QKD system, this work computationally explores the feasibility and potential of an OFDM-CVQKD protocol utilising these emerging THz sources. In Fig. 8(a), we simulate a wet open-air system ranging from single-carrier operation to $N=12$ subcarriers under different modulation variances. The results indicate that secure distances of up to 1.8 m (with $V_{mod}=1000$) and 1.55 m (with $V_{mod}=100$) are largely unaffected by the number of subcarriers, as the frequency lies within a THz atmospheric window. In this region, modulation noise is relatively weak and does not significantly degrade the SKR. It is also evident that higher modulation variance extends both the secure distance and the SKR. Therefore, achieving optimal performance requires balancing the robustness against modulation noise provided by low modulation variance in multi-subcarrier systems with the improved transmission capabilities associated with high modulation variance. In the scenario presented in Fig. 8(b), the optimal configuration (with 12 subcarriers in a dry open-air environment) yields secure links of over 3 m, making it well-suited for short-range wireless communications or local area networks between quantum computers. Considering that high-temperature superconducting coherent THz sources operate in cryostats, we evaluated their performance in a cryogenic or vacuum channel (see Fig. 9). With a 12 subcarrier configuration, the wireless link can extend up to roughly 26 km, presenting a promising solution for secure local area networks in quantum information processing as well as in space applications.

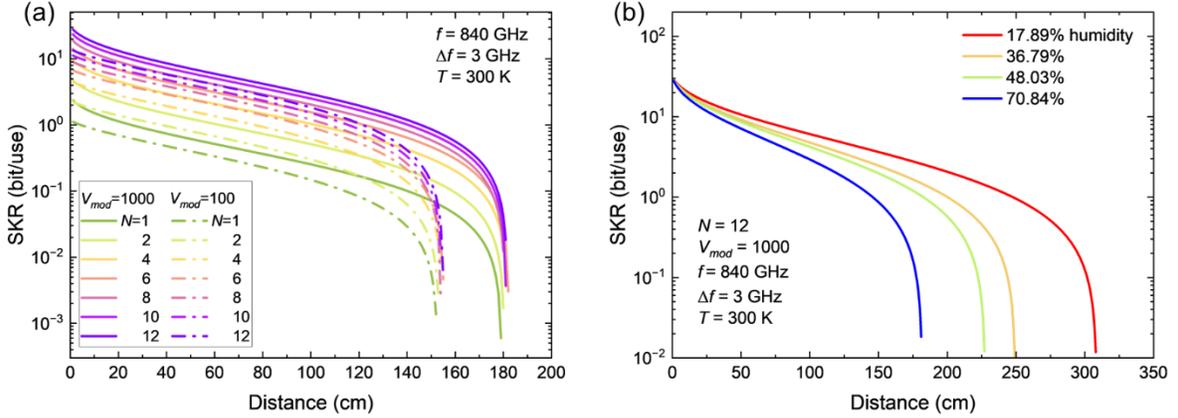

Fig. 8. (a) SKR as a function of transmission distance using homodyne detection at $f=840$ GHz, with $\Delta f = 3$ GHz and 70.84% relative humidity. Solid lines represent $V_{mod}=1000$ while dash-dotted lines indicate $V_{mod}=100$. (b) SKR as a function of transmission distance using homodyne detection at $f=840$ GHz, with $\Delta f = 3$ GHz under relative humidities of 17.89%, 36.79%, 48.03%, and 70.84% for a system with $N=12$ subcarriers and $V_{mod}=1000$.

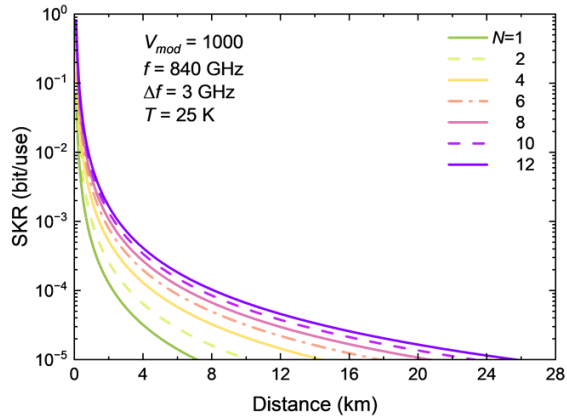

Fig. 9. SKR as a function of transmission distance in a vacuum channel at 25 K using homodyne detection with $f=840$ GHz, and $\Delta f = 3$ GHz. The modulation variance is $V_{mod}=1000$.

## VI. Conclusion

This study provides a thorough security analysis of the THz OFDM-CVQKD protocol in both open-air and inter-satellite environments, addressing key physical-layer challenges such as excess noise from I/Q imbalance and third-order intermodulation effects. Our findings demonstrate that utilising multiple subcarriers can enhance the secret key rate SKR to ~72 bits/channel use, a 94 times improvement over single-carrier implementations. However, this advantage is offset by increased modulation noise, particularly in high-subcarrier regimes, necessitating an optimised trade-off between subcarrier count, SKR, and secure transmission distance. In inter-satellite QKD, higher fundamental frequencies and an increased number of subcarriers enable secure transmission beyond 100 km. Unlike terrestrial channels, modulation noise remains the primary limitation rather than absorption losses, highlighting the need for high-fidelity THz modulation sources to fully exploit the OFDM architecture. While lower modulation variance enhances resilience against intermodulation noise, it demands precision-engineered THz sources for practical implementation. Our research highlights both the potential and limitations of the OFDM THz QKD protocol, with application-level considerations. We explored the feasibility of implementing an OFDM-CVQKD protocol in the THz regime, supported by recently developed superconducting THz sources. Using these properties as input to our protocol, we simulated secure key rates under various environmental conditions. Our results show that in wet or dry open-air channels, secure communication is possible over distances up to 3 m, while cryogenic or vacuum environments can support secure links up to ~26 km. These findings demonstrate that superconducting THz sources offer a viable platform for future chip-scale and wireless quantum communication systems. As THz device technology advances, particularly in chip-scale superconducting Josephson junction emitters, coherent THz modulators, and detectors, our findings provide a foundation for the next generation of high-speed, wireless quantum networks spanning terrestrial and space-based QKD infrastructures.

## Acknowledgements

This work, in part, was supported by the Royal Academy of Engineering (LTRF2223-19-138), the Royal Society of Edinburgh, and the Royal Society (RGS\R2\222168).